\documentclass[prd,aps,a4paper,superscriptaddress,twocolumn,nofootinbib]{revtex4-2}
\usepackage{dcolumn}
\usepackage{bm}
\usepackage{slashed}
\usepackage{amsmath}
\usepackage{amsthm}
\usepackage{latexsym}
\usepackage{amssymb}
\usepackage{mathrsfs}
\usepackage{amsfonts}
\usepackage{url}
\usepackage{graphicx}
\usepackage{hyperref}

\allowdisplaybreaks
\begin{document}

\title{Completing the real-space construction of the canonical harmonic coordinates of Kerr metric}

\author{Zizheng Xing}
\email[Zizheng Xing:~]{zizhengxing@sjtu.edu.cn}
\affiliation{School of Physics and Astronomy, Shanghai Jiao Tong University, Shanghai 200240, China}
\author{Xiaokai He}
\email[Xiaokai He:~]{sjyhexiaokai@hnfnu.edu.cn}
\affiliation{School of Mathematics and Statistics, and Hunan Provincial University Key Laboratory for Big Data Analysis and Application, Hunan First Normal University, Changsha 410205, China}
\author{Zhoujian Cao\footnote{corresponding author}}
\email[Zhoujian Cao:~]{zjcao@bnu.edu.cn}
\affiliation{School of Physics and Astronomy, Beijing Normal University, Beijing 100875, China}
\affiliation{School of Fundamental Physics and Mathematical Sciences, Hangzhou Institute for Advanced Study, University of Chinese Academy of Sciences, 
Hangzhou 310024, China}

\begin{abstract}
In this paper, we determine the exact canonical harmonic Kerr metric to all post-Minkowskian (PM) orders within the multipolar post-Minkowskian (MPM) formalism. 
The metric agrees with the previously constructed 4PM results.
Using a formal series ring, we prove that no homogeneous components of Laplacian operator appear in this metric at PM order $n\ge2$.
Together with the equality of the linear field with the canonical Kerr solution, this property allows an inductive construction of the canonical MPM metric where all gauge moments vanish.
The coordinate transformation linking the canonical harmonic Kerr metric to the Boyer-Lindquist metric is generated by the real part of a complex harmonic radial function. 
A recent independent construction by Sadu and Vanhove using momentum-space recursion and a residual homogeneous freedom classification yields the same result, providing a cross validation of the closed-form canonical harmonic Kerr metric.
\end{abstract}

\maketitle

\section{Introduction}

The multipolar post-Minkowskian (MPM) formalism~\cite{Blanchet.2024,Blanchet.1986} provides a systematic recursive construction for vacuum harmonic metrics outside isolated sources.
Its basic variable is the gothic metric perturbation
$h^{\mu\nu}_{\mathrm{MPM}} \equiv \sqrt{-g}\,g^{\mu\nu}-\eta^{\mu\nu}=\sum_{n=1}^\infty G^n h_{(n)}^{\mu\nu}$, which is constructed recursively as power series of the gravitational constant $G$.
The general solution is parametrized by six sets of symmetric and trace-free (STF) source moments:
two physical moments $\mathrm{I}_L, \mathrm{J}_L$ (mass-type and current-type respectively) and four gauge moments $\mathrm{W}_L,\mathrm{X}_L,\mathrm{Y}_L,\mathrm{Z}_L$.

The source moments may themselves depend on $G$, so the MPM solution has a two-layer dependence on $G$. 
The first layer is the expansion $\sum_{n=1}^{\infty} G^nh_{(n)}^{\mu\nu}$. 
The second layer is that $h_{(n)}^{\mu\nu}$ is parameterized by moments that depend on $G$.
At each PM iteration, the PM coefficients of source moments supply homogeneous solutions of the wave operator. The six source moments consequently parametrize the general harmonic vacuum solution, with different gauge moments corresponding to different coordinate conditions
(see Theorem~2 in \cite{Blanchet.2024}).

The system can also be parametrized isometrically by only two canonical moments $\mathrm{M}_L,\mathrm{S}_L$ where gauge moments vanish.
This greatly reduces the coordinate freedom and fixes the gauge in a canonical way, up to an arbitrary length scale $r_0$ introduced by finite-part (FP) regularization.
These two canonical moments do not parametrize the most general harmonic vacuum solution.
If the canonical moments and the source moments satisfy the relations reviewed in \cite{Blanchet.2024} (Eqs. (251-253)), the two metric functionals are related by a coordinate transformation and are isometric to each other.


In a previous work~\cite{kerr4PM}, we determined the canonical moments $\mathrm{M}_L$ and $\mathrm{S}_L$ of the Kerr spacetime and constructed its canonical harmonic metric up to fourth post-Minkowskian (4PM) order. The transformation to from the Boyer-Lindquist to the canonical harmonic metric is derived to 3PM there. 

After the work of~\cite{kerr4PM}, we derived the 4PM transformation and find that it can be reproduced from a simple closed-form real-part transformation.
The transformation associated with the complex function has been studied for a long time
\cite{Chou.1984,Abe.1987,Cook.1997,Jiang.2014.GRG}, and transformation associated with the real part was written explicitly in \cite{Ruiz.1986,Zang.2020}. 

By using a formal series ring that tracks the spin power and the STF order, we prove that the PM expansion of this metric has no homogeneous components of the Laplacian operator at any order $n\geq2$. 
The ordinary inverse Laplacian in real space reconstructs the metric from the 1PM coefficient which is exactly the 1PM canonical solution.
In this construction the FP regularization and harmonicity correction are not required, so the canonical harmonic metric is determined without the arbitrary length scale $r_0$. 
This shows that the canonical harmonic metric is unique and has minimal gauge content in the MPM sense.
 
If the imaginary part is introduced at 2PM, the homogeneous components appear alongside the particular solution generated by the direct inverse Laplacian.  
This implies that the real-part tranformation occupies a special position within the class of harmonic transformations \eqref{transU}. 
However, this real-part function exhibits infinite oscillations near the horizon, making the canonical harmonic Kerr coordinates not horizon-penetrating.

A recent work by Sadu and Vanhove \cite{sadu.2026} derived the same canonical harmonic coordinate to all orders by momentum-space recursion. 
The method of formal series ring in this paper and their classification of the residual homogeneous freedom provide an independent cross check for the closed-form canonical harmonic Kerr coordinates.

The paper is organized as follows. 
Section II briefly reviews the MPM formalism and the recursive algorithm.
Section III solves the harmonic coordinate transformations.
Section IV proves that the real-part transformation produces the canonical harmonic metric to all PM orders.
Section V provides a summary and discussion.
The absence of homogeneous components used in section IV is proved in Appendix A.

Throughout the paper, we explicitly retain $G$ and $c$, and use $\eta^{\mu\nu}=\{-1,1,1,1\}$.

\section{A review of the MPM formalism}

The MPM formalism uses the harmonic coordinate $x^\alpha=(ct,\boldsymbol{x})$ and uses $h^{\mu\nu}\equiv\sqrt{-g}\,g^{\mu\nu}-\eta^{\mu\nu}$ as the basic variable.
The equations for $h^{\mu\nu}$ are \cite{Blanchet.2024}
\begin{align}
    \Box h^{\alpha\beta}=\dfrac{16\pi G}{c^4} (-g)T^{\alpha\beta}+\Lambda^{\alpha\beta}_{\mathrm{harm}}(h),\label{relaxedEq}\\
    \partial_\beta h^{\alpha\beta}(ct,\boldsymbol{x})=0,\label{harmicity1}
\end{align}
which are equivalent to the Einstein field equation in the harmonic gauge.
Here $T^{\alpha\beta}$ stands for the energy-momentum tensor of matter.   $\Box=-\partial_0^2+\Delta$ is the flat d'Alembert operator. 
The harmonicity condition \eqref{harmicity1} is equivalent to
\begin{align}
    \nabla_{\mu}\nabla^\mu x^\alpha\equiv\dfrac{1}{\sqrt{-g}}\partial_\mu(\sqrt{-g}g^{\mu\nu}\partial_\nu x^\alpha)=0.
\end{align}
$\Lambda_{\mathrm{harm}}^{\alpha\beta}(h)$ is the pseudo energy-momentum tensor of the gravity itself in the wave equation [Eq. (54) in \cite{Blanchet.2024}].
Expanding the harmonic metric $h^{\mu\nu}=\sum_{n=1}^{\infty} G^n h^{\mu\nu}_{(n)}$ in powers of the gravitational constant $G$, one can decompose the vacuum equations into
\begin{subequations}
\begin{align}
    \Box h^{\alpha\beta}_{(n)}&=\Lambda_{\mathrm{harm}(n)}^{\alpha\beta}(h_{(1)},\cdots,h_{(n-1)}),\\
    \partial_\beta h^{\alpha\beta}_{(n)}&=0.
\end{align}
\end{subequations}
Here $\Lambda_{\mathrm{harm}(n)}^{\alpha\beta}$ is generated by expanding $\Lambda_{\mathrm{harm}}^{\alpha\beta}(h)$.

Ref.~\cite{Blanchet.2024} gives the multipolar post-Minkowskian algorithm for constructing the general harmonic vacuum solutions $h_{\mathrm{MPM}}^{\mu\nu}=\sum_{n=1}^{\infty} G^n h_{(n)}^{\mu\nu}$. 
The first-order solution $h_{(1)}^{\mu\nu}$ is parameterized by $\{\mathrm{I}_L,\mathrm{J}_L,\mathrm{W}_L,\mathrm{X}_L,\mathrm{Y}_L,\mathrm{Z}_L\}$ or isometrically by $\{\mathrm{M}_L,\mathrm{S}_L\}$ [Eq. (64-66) in \cite{Blanchet.2024}]. 
For $n\ge2$, the functional $h_{(n)}^{\mu\nu}$ is defined recursively by the lower order solutions
\begin{subequations}\label{MPMalgorithm}
\begin{align}
    &h_{(n)}^{\mu\nu}=u_{(n)}^{\mu\nu}+v_{(n)}^{\mu\nu},\\
    &u_{(n)}^{\mu\nu}=\underset{B=0}{\mathrm{FP}}\Box_{\mathrm{ret}}^{-1}\Bigl[\Bigl(\frac{r}{r_0}\Bigr)^B \Lambda_{\mathrm{harm}(n)}^{\mu\nu}(h_{(1)},\cdots,h_{(n-1)})\Bigr],\\
    &(\Box_{\mathrm{ret}}^{-1}f)_{(t,\bm{x})}=-\dfrac{1}{4\pi}\int \dfrac{\mathrm{d}^3\bm{y}}{|\bm{x-\bm{y}}|}f(t-|\bm{x-\bm{y}}|/c,\bm{y}),\\
    &v_{(n)}^{\mu\nu}=\mathcal{V}^{\mu\nu}(w_{(n)}^\alpha),\quad w_{(n)}^{\alpha}=\partial_\beta u_{(n)}^{\alpha\beta}.
\end{align}
\end{subequations}
Here $r_0$ is an arbitrary constant length scale.
The STF notation, FP regularization and the harmonicity correction $\mathcal{V}^{\mu\nu}$ are defined in detail in footnote 23 and Eqs. (70-75) and Eq. (76-79) in the review~\cite{Blanchet.2024}. 



For stationary configurations the d'Alembert operator reduces to the Laplacian operator. 
The Laplacian of functions $\dfrac{n_{\langle L \rangle}}{r^{k}}$ is given as
\begin{align}\label{Laplacian}
    \Delta\left(\dfrac{n_{\langle L \rangle}}{r^{k}}\right)=(k+l)(k-l-1)\dfrac{n_{\langle L \rangle}}{r^{k+2}}.
\end{align}
Here the subscript $L$ is the shorthand of spatial indices $i_1i_2\cdots i_l$. $n_{\left\langle L\right\rangle}$ is the STF part of the spatial tensor $n_{i_1}n_{i_2}\cdots n_{i_l}$. 

The inverse of $\dfrac{n_{\langle L \rangle}}{r^{k+2}}$ is given as (Eq. (3.9) in \cite{Blanchet.1986})
\begin{align}\label{inverseLaplacian}
    \Delta^{-1}\left(\dfrac{n_{\langle L \rangle}}{r^{k+2}}\right)=\dfrac{n_{\langle L \rangle}}{(k+l)(k-l-1)r^{k}},
\end{align}
if the denominators do not vanish. 
If the denominator of \eqref{inverseLaplacian} vanishes, the FP operation $\mathrm{FP}_{B=0}$ is required.

For a single term $d(\boldsymbol{x})$ in a series $f(\boldsymbol{x})$, if the FP operator is required to compute $\Delta^{-1}d(\boldsymbol{x})$, we call this single term $d(\boldsymbol{x})$ dangerous in the series $f(\boldsymbol{x})$.

We see that $\Delta \dfrac{n_{\langle L \rangle}}{r^{l+1}}=0$ and call the functions $\dfrac{n_{\langle L \rangle}}{r^{l+1}}$ the kernel or the homogeneous components of the Laplacian operator. 
If a series $f(\boldsymbol{x})=\sum \dfrac{n_{\langle L \rangle}}{r^{k}}$ has some homogeneous components, the inverse operator cannot reproduce these homogeneous components from zero, therefore $f\neq\Delta^{-1}(\Delta f)$.
Here $\Delta^{-1}$ is well-defined through \eqref{inverseLaplacian} since there are no dangerous terms in $(\Delta f)$ due to the coefficient in \eqref{Laplacian}.
If the series has no homogeneous components, we derive $f=\Delta^{-1}(\Delta f)$. 

The canonical harmonic Kerr metric $h_{\mathrm{can}}^{\mu\nu}$ has no gauge moments at all PM orders. 
The canonical Kerr moments are given by \cite{kerr4PM}
\begin{subequations}\label{MLSL}
\begin{align}
    \mathrm{M}_L&=\left\{\begin{aligned}
        &(-1)^{l/2}M a^l (\boldsymbol{e}_z)_{\langle L\rangle},&&\text{$l$ even},\\
        &0,&&\text{$l$ odd},
    \end{aligned}\right.\\
    \mathrm{S}_L&=\left\{\begin{aligned}
        &(-1)^{(l-1)/2}\dfrac{(l+1)c}{2l}M a^l (\boldsymbol{e}_z)_{\langle L\rangle},&&\text{$l$ odd},\\
        &0,&&\text{$l$ even}.
    \end{aligned}\right.
\end{align}
\end{subequations}
With these moments $h_{\mathrm{can}(n)}^{\mu\nu}$ can be constructed in principle to any PM order. 
The 4PM order results can be found in \cite{kerr4PM}.
The 1PM metric satisfies $\Delta h_{\mathrm{can}(1)}^{\mu\nu}=0$, therefore $h_{\mathrm{can}(1)}^{\mu\nu}$ has only homogeneous components and $h_{\mathrm{can}(1)}^{\mu\nu}\neq\Delta^{-1}(\Delta h_{\mathrm{can}(1)}^{\mu\nu})$.

\section{Harmonic coordinates transformation}

The Kerr metric is parameterized by $m\equiv GM/c^2$ and $a$.  We take both $m$ and $a$ to have dimensions of length. The spin parameter $a$ is held fixed and is not expanded in $G$.
We consider the following type of coordinate transformations
\begin{align}
    \left\{
    \begin{aligned}
    T&=t,\\
    X+\mathrm{i}Y&=U(r_{\mathrm{BL}})\,\sin\theta_{\mathrm{BL}}\,\mathrm{e}^{\mathrm{i}\varphi_{\mathrm{BL}}},\\
    Z&=(r_{\mathrm{BL}}-m)\cos\theta_{\mathrm{BL}}.
    \end{aligned}\right.\label{transU}
\end{align}
Here  $\{x^0=ct,r_{\mathrm{BL}},\theta_{\mathrm{BL}},\varphi_{\mathrm{BL}}\}$ are the Boyer-Lindquist coordinates \cite{Boyer.Lindquist.1967}. 
The complex function $U(r_{\mathrm{BL}})$ has the dimension of length. The harmonicity condition on $X^{\alpha}=\{ct,\boldsymbol{X}\}$ reduces to
\begin{align}
    \dfrac{\mathrm{d}}{\mathrm{d}r_{\mathrm{BL}}}\left(\varDelta \dfrac{\mathrm{d}}{\mathrm{d}r_{\mathrm{BL}}}U\right)+\left(\dfrac{a^2}{\varDelta}-2\right)U=0.\label{DU}
\end{align}
Here $\varDelta=(r_{\mathrm{BL}}^2-2mr_{\mathrm{BL}}+a^2)$. A complex solution of \eqref{DU} is given by $\chi(r_{\mathrm{BL}})$
\begin{align}
    &\chi(r_{\mathrm{BL}})=(r_{\mathrm{BL}}-m+\mathrm{i}a)e^{\mathrm{i}\psi(r_{\mathrm{BL}})},\\
    &\psi(r_{\mathrm{BL}})=\int\dfrac{a\,\mathrm{d}r_{\mathrm{BL}}}{\varDelta(r_{\mathrm{BL}})}\nonumber\\
    &=\left\{\begin{aligned}
        &-\frac{a}{\sqrt{m^2-a^2}} \mathrm{arctanh}\left(\frac{\sqrt{m^2-a^2}}{r_{\mathrm{BL}}-m}\right),\,a^2<m^2,\label{psi}\\
        &-\frac{a}{r_{\mathrm{BL}}-m},\quad a^2=m^2,\\
        &-\frac{a}{\sqrt{a^2-m^2}} \mathrm{arctan}\left(\frac{\sqrt{a^2-m^2}}{r_{\mathrm{BL}}-m}\right),\,a^2>m^2,
    \end{aligned}\right.
\end{align}
This radial function $\chi$ recovers the transformations known in \cite{Chou.1984,Abe.1987,Cook.1997,Jiang.2014.GRG}.

The real part $H=\mathrm{Re}\chi$ and the imaginary part $\mathrm{Im}\chi$ are
\begin{align}
    H(r_{\mathrm{BL}})&=(r_{\mathrm{BL}}-m)\cos\psi-a\sin\psi,\label{H}\\
    \mathrm{Im}\chi&=(r_{\mathrm{BL}}-m)\sin\psi+a\cos\psi.
\end{align}

The general solution of Eq. \eqref{DU} is a linear combination of $H$ and $\mathrm{Im}\chi$
\begin{align}
    U(r_{\mathrm{BL}})=c_1H(r_{\mathrm{BL}})+c_2\mathrm{Im}\chi(r_{\mathrm{BL}}).\label{U}
\end{align}
Here $c_1$ and $c_2$ are two complex constants. Eq. \eqref{transU} is not the most general harmonic coordinate; see \cite{Ruiz.1986,LiuQuanhui.1998} for the general case. 
The far field expansions of $H$ and $H$ are given by
\begin{align}
    H&=\rho+\dfrac{a^2}{2\rho}-\dfrac{a^4}{8\rho^3}+\dfrac{a^2(9a^4-8m^4)}{144\rho^5}+O\left(\rho^{-7}\right),\label{Frho}\\
    \mathrm{Im}\chi&=-\dfrac{a m^2}{3\rho^2}+\dfrac{a m^2(7a^2-6m^2)}{30\rho^4}+O\left(\rho^{-6}\right).\label{Hrho}
\end{align}
Here $\rho\equiv r_{\mathrm{BL}}-m$. To guarantee the far field asymptotic properties of the transformation, $c_1c_1^*=1$, or equivalently $c_1=1$ up to a constant rotation. 
Any deviation $c_1\neq1$ will introduce constants independent of $(X,Y,Z)$ in $\sqrt{-g}g^{\mu\nu}-\eta^{\mu\nu}$, whereas the MPM algorithm requires such terms to vanish at spatial infinity. 
Therefore the harmonic metric with $c_1\neq1$ cannot be reproduced through the MPM algorithm.

The harmonic transformation of the real part \eqref{H} is given as
\begin{align}
    &\left\{
    \begin{aligned}
    T&=t,\\
    X&=H(r_{\mathrm{BL}})\,\sin\theta_{\mathrm{BL}}\,\cos\varphi_{\mathrm{BL}},\\
    Y&=H(r_{\mathrm{BL}})\,\sin\theta_{\mathrm{BL}}\,\sin\varphi_{\mathrm{BL}},\\
    Z&=(r_{\mathrm{BL}}-m)\cos\theta_{\mathrm{BL}}.
    \end{aligned}\right.\label{trans}
\end{align}
This real part coordinate transformation was written explicitly in \cite{Ruiz.1986,Zang.2020}.

We denote $h_{H}^{\mu\nu}$ and $h_{H(n)}^{\mu\nu}$ as the harmonic metric generated by transformation \eqref{trans} and the coefficient of $G^n$ in its expansion.
We have checked that this metric has no deviation from the canonical harmonic Kerr metric at 4PM order
\begin{align}
    h_{H(i)}^{\mu\nu}=h_{\mathrm{can}(i)}^{\mu\nu},\quad 1\le i\le4.
\end{align}
Since $h_{H}^{\mu\nu}$ is an exact vacuum solution of  \eqref{relaxedEq}, one derives
\begin{align}
    \Lambda_{\mathrm{harm}}^{\mu\nu}(h_{H})&=\Delta h_{H}^{\mu\nu},\quad
    \Lambda_{\mathrm{harm}(n)}^{\mu\nu}(h_{H})=\Delta h_{H(n)}^{\mu\nu}.
\end{align} 
According to the transformation \eqref{trans}, 
the $1/r$ expansion of $h_H^{\mu\nu}$ takes the form
\begin{align}
    h_{H}^{\mu\nu}&=Gh_{\mathrm{can}(1)}^{\mu\nu}+\sum_{n\ge2}\sum_{p}\sum_{l}\dfrac{G^n}{c^{2n}}\dfrac{M^na^pn_{\langle L\rangle}}{r^{n+p}}h_{n,p,L}^{\mu\nu}.\label{hHsum}
\end{align}
Here $r=|\boldsymbol{X}|,\boldsymbol{n}=\boldsymbol{X}/r$. $h_{n,p,L}^{\mu\nu}$ are dimensionless numerical tensors independent of $(r,\boldsymbol{n})$. 

Using a formal series ring to track the angular power $l$ and the radial power $n+p$, we prove in Appendix A that for $n\ge 2$, $h_{H(n)}^{\mu\nu}$ has no homogeneous components which require $l=n+p-1$. 
This property guarantees that 
\begin{align}
    h_{H(n)}^{\mu\nu}=\Delta^{-1}(\Delta h_{H(n)}^{\mu\nu}),\quad n\ge2.\label{NoHomogeneous}
\end{align}

\section{Proof of the canonical harmonic Kerr metric}
In this section we prove that the harmonic metric $h_{H}^{\mu\nu}$ generated by \eqref{trans} is the canonical harmonic Kerr metric $h_{\mathrm{can}}^{\mu\nu}$. 

Assuming that for $n\ge2$ we have $h_{H(i)}^{\mu\nu}=h_{\mathrm{can}(i)}^{\mu\nu}$ for all $1\le i\le n-1$, we then prove that $h_{H(n)}^{\mu\nu}=h_{\mathrm{can}(n)}^{\mu\nu}$. The source $\Lambda_{\mathrm{harm}(n)}^{\mu\nu}$ in \eqref{MPMalgorithm} is a functional of $(h_{\mathrm{can}(1)},\cdots,h_{\mathrm{can}(n-1)})$, therefore by assumption we have 
\begin{align}
    \Lambda_{\mathrm{harm}(n)}^{\mu\nu}=\Lambda_{\mathrm{harm}(n)}^{\mu\nu}(h_{H(1)},\cdots,h_{H(n-1)}).
\end{align} 
The right hand side is also the $G^n$ coefficient of the full source $\Lambda_{\mathrm{harm}}^{\mu\nu}(h_{H})=\Delta h_{H}^{\mu\nu}$, therefore $\Lambda_{\mathrm{harm}(n)}^{\mu\nu}=\Delta h_{H(n)}^{\mu\nu}$. 

Based on \eqref{NoHomogeneous} (proven in Appendix A), the direct inverse is simply  $u_{(n)}^{\mu\nu}=\mathrm{FP}\Delta^{-1}(\Lambda_{\mathrm{harm}(n)}^{\mu\nu})=\Delta^{-1}(\Delta h_{H(n)}^{\mu\nu})=h_{H(n)}^{\mu\nu}$. Since $u_{(n)}^{\mu\nu}=h_{H(n)}^{\mu\nu}$ is harmonic, the harmonicity correction $v^{\alpha\beta}_{(n)}$ in \eqref{MPMalgorithm} is not required, hence we derive $h_{\mathrm{can}(n)}^{\mu\nu}=h_{H(n)}^{\mu\nu}$ under the inductive assumption. 
Since we have checked that $h_{\mathrm{can}(1)}^{\mu\nu}=h_{H(1)}^{\mu\nu}$, we prove by induction that the harmonic metric $h_{H}^{\mu\nu}$ generated by \eqref{trans} is the canonical harmonic Kerr metric $h_{\mathrm{can}}^{\mu\nu}$.

\section{Summary and Discussion}
In this paper, we prove that $h_{H}^{\mu\nu}$ generated by \eqref{trans} is the exact canonical harmonic Kerr metric to all PM orders. 

It is straightforward to verify from Appendix A that $h_{H}^{00}$ and $h_{H}^{ij}$ are even in $a$, while $h_{H}^{0i}$ is odd in $a$. Furthermore $h_H^{ij}$ is even in $m$. 
The proof in Section IV also implies that FP regularization and harmonicity correction are not required. 
These properties all agree with the assumptions and observations in \cite{kerr4PM}.

Ref.~\cite{Zang.2020} pointed out that the real part function $H(r_{\mathrm{BL}})$ exhibits shocks and oscillations near the event horizon $r_+ = m + \sqrt{m^2-a^2}$. In fact, near the horizon $r_+$, the function $H(r_{\mathrm{BL}})$ takes the asymptotic form
\begin{align}
    H(r_{\mathrm{BL}})\simeq m \cos\left[\dfrac{a}{2b}\ln\left(\dfrac{r_{\mathrm{BL}}-r_+}{2b}\right)+\arctan\dfrac{a}{b}\right],
\end{align}
for $a^2<m^2$ where $b=\sqrt{m^2-a^2}$. Therefore, unlike the metrics in \cite{Cook.1997,Jiang.2014.GRG,Zang.2020}, the canonical harmonic metric is not horizon penetrating.

Other harmonic transformations \eqref{U} do not produce the canonical MPM gauge and have moments more complex than those in\eqref{MLSL}.
For example the Jiang-Lin metric [Eq. (15) in \cite{Jiang.2014.GRG}] is produced by the transformation $U=\chi(r_{\mathrm{BL}})$, and has the similar expansion form as \eqref{hHsum}, but deviates from the canonical metric at 2PM since $\mathrm{Im}\chi$ starts at 2PM. 
Therefore one derives $h_{\mathrm{JL}[1]}^{\mu\nu}=h_{\mathrm{can}(1)}^{\mu\nu}$ and $\Delta h_{\mathrm{JL}[2]}^{\mu\nu}=\Lambda_{\mathrm{harm}(2)}^{\mu\nu}(h_{\mathrm{JL}})=\Lambda_{\mathrm{harm}(2)}^{\mu\nu}(h_{\mathrm{can}})$. The solution $h_{\mathrm{JL}[2]}^{\mu\nu}$ is the particular solution $h_{\mathrm{can}(2)}^{\mu\nu}$ plus a homogeneous term $h_{(1)}^{\mu\nu}[\mathrm{Z}_L]$, which can be contributed by the gauge moments $\mathrm{Z}_L\sim O(G)$ at 1PM order [Eq. (62) in \cite{kerr4PM}].

The metrics of harmonic transformations \eqref{U} exhibits similar expansion as \eqref{hHsum}. Suppose that $c_2\neq0$, and let $n\geq2$ be the first PM order at which $h_{U[n]}^{\mu\nu}\neq h_{\mathrm{can}(n)}^{\mu\nu}$
Since $\Lambda_{\mathrm{harm}(n)}^{\mu\nu}(h_{U})=\Lambda_{\mathrm{harm}(n)}^{\mu\nu}(h_{\mathrm{can}})$, we derive $\Delta (h_{U[n]}^{\mu\nu}-h_{\mathrm{can}(n)}^{\mu\nu})=0$. Therefore the homogeneous component appears for these metrics. 

\appendix
\section{Absence of Homogeneous Components in \texorpdfstring{$h_{H(n)}^{\mu\nu}$}{} when \texorpdfstring{$n\ge2$}{}}

In this Appendix, we prove that $h_{H(n)}^{\mu\nu}$ has no homogeneous components for $n\ge2$.

The metric $h_{H}^{\mu\nu}$ generated by \eqref{trans} can be written as
\begin{subequations}\label{hH}
\begin{align}
    h^{00}_H&=1+A,\\
    h^{01}_H&=n_2\dfrac{a}{r}C,\quad h^{02}_H=-n_1\dfrac{a}{r}C,\quad h^{03}_H=0,\\
    h^{11}_H&=-1+A_{\perp}+(A_{\parallel}-A_{\perp})\dfrac{n_1^2}{1-\zeta^2},\\
    h^{22}_H&=-1+A_{\perp}+(A_{\parallel}-A_{\perp})\dfrac{n_2^2}{1-\zeta^2},\\
    h^{12}_H&=(A_{\parallel}-A_{\perp})\dfrac{n_1n_2}{1-\zeta^2},\\
    h^{13}_H&=n_1\zeta B,\quad h^{23}_H=n_2\zeta B,\\
    h^{33}_H&=-1+A_{zz}.
\end{align}
\end{subequations}
Here $\zeta=n_3=Z/r$, and $A,C,A_{\parallel},A_{\perp},A_{zz},B$ are functions of $(r,\boldsymbol{n})$. 
These functions are constructed from simpler functions of $(r,\zeta^2)$
\begin{subequations}\label{functions}
\begin{align}
    x&=\dfrac{\rho}{r},\quad \rho=r_{\mathrm{BL}}-m,\quad s=\dfrac{a^2}{r^2},\quad w=\dfrac{m}{r},\\
    L&=\dfrac{H}{\rho},\quad P=\dfrac{\mathrm{d}H}{\mathrm{d}\rho},\\
    \mathcal{E}&=L^2\zeta^2+\dfrac{P}{L}(1-\zeta^2),\quad
    \mathcal{D}=x^2-w^2+s,\\
    A&=\dfrac{1}{\mathcal{E}}\left(\dfrac{s(1-\zeta^2)}{x^2L^2}-\dfrac{((x+w)^2+s)^2}{\mathcal{D}}\right),\\
    C&=\dfrac{2w(x+w)}{\mathcal{E}\mathcal{D}},\\
    A_{\parallel}&=\dfrac{1}{\mathcal{E}}\left(\dfrac{\mathcal{D}P^2(1-\zeta^2)}{x^2L^2}+L^2\zeta^2\right),\\
    A_{\perp}&=\dfrac{1}{\mathcal{E}}\left(x^2L^2-\dfrac{s(1-\zeta^2)}{\mathcal{D}}\right),\\
    A_{zz}&=\dfrac{1}{\mathcal{E}}\left(\dfrac{\mathcal{D}\zeta^2}{x^2}+\dfrac{1-\zeta^2}{L^2}\right),\\
    B&=\dfrac{1}{\mathcal{E}}\left(\dfrac{\mathcal{D}P}{x^2L}-1\right).
\end{align}
\end{subequations}
From \eqref{trans}, we have $X^2+Y^2=H^2(1-Z^2/\rho^2)$, which establishes the implicit relation among $x,L$ and $\zeta$
\begin{align}
    x^2=\dfrac{1}{L^2}+\zeta^2\left(1-\dfrac{1}{L^2}\right),\label{xL1}
\end{align}
Equations \eqref{psi} and \eqref{H} can be expanded as
\begin{align}
    \psi&=-a\sum_{i\ge0}\dfrac{(m^2-a^2)^i}{(2i+1)\rho^{2i+1}},\\
    \dfrac{H}{\rho}=L&=1+\sum_{\alpha\ge1,\beta\ge0}\left(\dfrac{a}{\rho}\right)^{2\alpha}\left(\dfrac{m}{\rho}\right)^{2\beta}\mathcal{F}_1.\label{xL2}
\end{align}
Here $\mathcal{F}$ represents coefficients independent of $(m,a)$. The summation range $\alpha\ge1$ arises from the overall factor $a$ in the $\psi$ expansion. 
With equations \eqref{xL1} and \eqref{xL2}, one can solve $x,L$ as formal power series of the form $\sum_{k,p,l} \dfrac{m^{k-p}a^p\zeta^l}{r^k}$.

$P,\dfrac{1}{L^2}$ can be expanded in the same form as \eqref{xL2} with different coefficients. Therefore, $x^2$ takes the form
\begin{align}
    x^2=1+\sum_{\alpha\ge1,\beta\ge0}\left(\dfrac{a}{\rho}\right)^{2\alpha}\left(\dfrac{m}{\rho}\right)^{2\beta}\zeta^{2\gamma}\mathcal{F}_2.
\end{align}
Here $\gamma$ is $0$ or $1$. Since every $\zeta^2$ in $x^2$ is accompanied by at least one $a^2$, the expansion of $x$ in terms of $1/\rho$ takes the form
\begin{align}
    \dfrac{\rho}{r(\rho)}=1+\sum_{(k \text{ even})}\sum_{(p \text{ even})}\sum_{(l  \text{ even})}^p \dfrac{m^{k-p}a^p\zeta^l}{\rho^k}\mathcal{F}_3.
\end{align}
Solving $\rho(r)$ from this expansion reveals that the series of $x$ in terms of $1/r$ has the same form
\begin{align}
    x=1+\sum_{(k \text{ even})}\sum_{(p \text{ even})}\sum_{(l\text{ even})}^p \dfrac{m^{k-p}a^p\zeta^l}{r^k}\mathcal{F}_4.
\end{align}

Similarly, the series $L$ and $P$ take the same form except for the summation range $l\le p-2$, which arises from the condition $\alpha\ge1$ in \eqref{xL2}.

$x,L,P$ are three basic building blocks which produce other series by addition, multiplication and division. 
We introduce a set $\mathcal{J}$ containing formal series of the form $\sum\dfrac{m^{k-p}a^p\zeta^l}{r^k}$ where $l\le p$ and $l,p$ are even. 
This set is closed under addition and multiplication, and is closed under division by elements with nonzero constant term. $\mathcal{J}$ constitutes a formal series ring \cite{Algebra}.

It is straightforward to verify that $x,L,P,\mathcal{E},\mathcal{D},A,C,A_{\perp},B$ are elements of $\mathcal{J}$. 

For $A_{\parallel}$ and $A_{zz}$, the factor $1-\zeta^2$ is multiplied without being accompanied by $s=a^2/r^2$. 
Consequently, these two series satisfy the bound $l \le p+2$.

We have the identity 
\begin{align}
    A_{\parallel}-A_{\perp}=(1-\zeta^2)\dfrac{1}{\mathcal{E}}\left(\dfrac{\mathcal{D}P^2}{x^2L^2}+\dfrac{s}{\mathcal{D}}-1\right),    
\end{align}
therefore the denominators $1-\zeta^2$ in $h_H^{ij}$ are canceled and do not break the series structure. 

With the series \eqref{functions} in hand, we construct $h_H^{\mu\nu}$ through \eqref{hH} and derive that $h_H^{\mu\nu}$ takes the form $\sum \dfrac{m^{n}a^pn_{\langle L\rangle}}{r^{n+p}}$.

For a single term $\dfrac{m^{n}a^pn_{\langle L\rangle}}{r^{n+p}}$ in the series $h_{H}^{00}$, we have $l\le p$. For $n\ge2$, it follows that $l<n+p-1$. 
Therefore there are no homogeneous components which require $l=n+p-1$ in $h_{H(n)}^{00}$ when $n\ge 2$.

For a single term $\dfrac{m^{n}a^{p+1}n_{\langle L+1\rangle}}{r^{n+p+1}}$ in the series $h_{H}^{0i}$, we have $l\le p$. For $n\ge2$, it follows that $l+1<n+p$. 
Therefore there are no homogeneous components which require $l+1=n+p$ in $h_{H(n)}^{0i}$ when $n\ge 2$.

$h_{H}^{ij}$ is even in $m$, since $m$ (or $w$) always appears squared in the space kernels $A_{\parallel},A_{\perp},A_{zz},B$. 
For a single term $\dfrac{m^{n}a^{p}n_{\langle L\rangle}}{r^{n+p}}$ in the series $h_{H}^{ij}$, the indices $l,n,p$ are even.
Therefore there are no homogeneous components which require $l=n+p-1$ in $h_{H(n)}^{ij}$ when $n\ge 2$, since $n+p-1$ is odd. 

For a brief check, we list the series up to $O(r^{-8})$ ($h_{\mathrm{can}(1)}^{00}$ and $h_{\mathrm{can}(1)}^{0i}$ can be found in \cite{kerr4PM}, while $h_{\mathrm{can}(1)}^{ij}=0$)
\begin{widetext}
\begin{subequations}
\begin{align}
x&=1
+\frac{a^{2}(\zeta^{2}-1)}{2r^{2}}
-\frac{a^{4}(\zeta^{2}-1)(5\zeta^{2}-1)}{8r^{4}}
+\frac{a^{2}(\zeta^{2}-1)
 \left[9a^{4}(21\zeta^{4}-14\zeta^{2}+1)-8m^{4}\right]}
 {144r^{6}}
+O(r^{-8}),\\
L&=1
+\frac{a^{2}}{2r^{2}}
-\frac{a^{4}(4\zeta^{2}-3)}{8r^{4}}
+\frac{a^{2}
 \left[9a^{4}(16\zeta^{4}-20\zeta^{2}+5)-8m^{4}\right]}
 {144r^{6}}
+O(r^{-8}),\\
P&=1
-\frac{a^{2}}{2r^{2}}
+\frac{a^{4}(4\zeta^{2}-1)}{8r^{4}}
-\frac{a^{2}
 \left[9a^{4}(16\zeta^{4}-12\zeta^{2}+1)-40m^{4}\right]}
 {144r^{6}}
+O(r^{-8}),\\
\mathcal{E}&=1
+\frac{a^{2}(2\zeta^{2}-1)}{r^{2}}
-\frac{2a^{4}\zeta^{2}(\zeta^{2}-1)}{r^{4}}
+\frac{a^{2}
 \left[
 9a^{4}(4\zeta^{6}-6\zeta^{4}+2\zeta^{2})
 -4m^{4}\zeta^{2}+3m^{4}
 \right]}
 {9r^{6}}
+O(r^{-8}),\\
\mathcal{D}&=1
+\frac{a^{2}\zeta^{2}-m^{2}}{r^{2}}
-\frac{a^{4}\zeta^{2}(\zeta^{2}-1)}{r^{4}}
+\frac{a^{2}(\zeta^{2}-1)
 \left[9a^{4}(2\zeta^{4}-\zeta^{2})-m^{4}\right]}
 {9r^{6}}
+O(r^{-8}),\\
A&=-1
-\frac{4m}{r}
-\frac{7m^{2}}{r^{2}}
+\frac{2m\left(3a^{2}\zeta^{2}-a^{2}-4m^{2}\right)}{r^{3}}
+\frac{m^{2}\left(14a^{2}\zeta^{2}-3a^{2}-8m^{2}\right)}{r^{4}}
\notag\\
&\quad
-\frac{m\left(
35a^{4}\zeta^{4}-30a^{4}\zeta^{2}+3a^{4}
-40a^{2}m^{2}\zeta^{2}+8a^{2}m^{2}+16m^{4}
\right)}{2r^{5}}
\notag\\
&\quad
-\frac{m^{2}\left(
378a^{4}\zeta^{4}-270a^{4}\zeta^{2}+27a^{4}
-213a^{2}m^{2}\zeta^{2}+34a^{2}m^{2}+72m^{4}
\right)}{9r^{6}}
\notag\\
&\quad
+\frac{m}{36r^{7}}\left(
2079a^{6}\zeta^{6}-2835a^{6}\zeta^{4}+945a^{6}\zeta^{2}-45a^{6}
\right.\notag\\
&\qquad\left.
-2268a^{4}m^{2}\zeta^{4}+1512a^{4}m^{2}\zeta^{2}-108a^{4}m^{2}
+952a^{2}m^{4}\zeta^{2}-104a^{2}m^{4}-288m^{6}
\right)
+O(r^{-8}),\\
C&=\frac{2m}{r}
+\frac{2m^{2}}{r^{2}}
-\frac{m\left(5a^{2}\zeta^{2}-a^{2}-2m^{2}\right)}{r^{3}}
-\frac{2m^{2}\left(3a^{2}\zeta^{2}-a^{2}-m^{2}\right)}{r^{4}}
\notag\\
&\quad
+\frac{m\left(
63a^{4}\zeta^{4}-42a^{4}\zeta^{2}+3a^{4}
-28a^{2}m^{2}\zeta^{2}+4a^{2}m^{2}+8m^{4}
\right)}{4r^{5}}
\notag\\
&\quad
+\frac{2m^{2}\left(
10a^{4}\zeta^{4}-8a^{4}\zeta^{2}+a^{4}
-4a^{2}m^{2}\zeta^{2}+a^{2}m^{2}+m^{4}
\right)}{r^{6}}
\notag\\
&\quad
-\frac{m}{72r^{7}}\left(
3861a^{6}\zeta^{6}-4455a^{6}\zeta^{4}+1215a^{6}\zeta^{2}-45a^{6}
\right.\notag\\
&\qquad\left.
-1782a^{4}m^{2}\zeta^{4}+972a^{4}m^{2}\zeta^{2}-54a^{4}m^{2}
+576a^{2}m^{4}\zeta^{2}-16a^{2}m^{4}-144m^{6}
\right)
+O(r^{-8}),\\
A_{\perp}&=1
+\frac{a^{2}m^{2}(\zeta^{2}-1)}{r^{4}}
-\frac{a^{2}m^{2}(\zeta^{2}-1)
 \left(12a^{2}\zeta^{2}-3a^{2}-4m^{2}\right)}
 {3r^{6}}
+O(r^{-8}),\\
B&=-\frac{m^{2}}{r^{2}}
+\frac{a^{2}m^{2}(3\zeta^{2}-1)}{r^{4}}
-\frac{a^{2}m^{2}
 \left(30a^{2}\zeta^{4}-24a^{2}\zeta^{2}+3a^{2}-m^{2}\right)}
 {3r^{6}}
+O(r^{-8}),\\
A_{\parallel}&=1
+\frac{m^{2}(\zeta^{2}-1)}{r^{2}}
-\frac{3a^{2}m^{2}\zeta^{2}(\zeta^{2}-1)}{r^{4}}
+\frac{a^{2}m^{2}(\zeta^{2}-1)
 \left(30a^{2}\zeta^{4}-12a^{2}\zeta^{2}-m^{2}\right)}
 {3r^{6}}
+O(r^{-8}),\\
A_{zz}&=1
-\frac{m^{2}\zeta^{2}}{r^{2}}
+\frac{a^{2}m^{2}\zeta^{2}(3\zeta^{2}-2)}{r^{4}}
-\frac{a^{2}m^{2}
 \left[
 90a^{2}\zeta^{6}-108a^{2}\zeta^{4}+27a^{2}\zeta^{2}
 -3m^{2}\zeta^{2}+2m^{2}
 \right]}
 {9r^{6}}
+O(r^{-8}),\\
h_H^{00}
&=Gh_{\mathrm{can}(1)}^{00}
-\frac{7m^{2}}{r^{2}}
-\frac{8m^{3}}{r^{3}}
+\frac{m^{2}\left(14a^{2}\zeta^{2}-3a^{2}-8m^{2}\right)}{r^{4}}
+\frac{4m^{3}\left[a^{2}(5\zeta^{2}-1)-2m^{2}\right]}{r^{5}}
\notag\\
&\quad
-\frac{m^{2}\left(
378a^{4}\zeta^{4}-270a^{4}\zeta^{2}+27a^{4}
-213a^{2}m^{2}\zeta^{2}+34a^{2}m^{2}+72m^{4}
\right)}{9r^{6}}
\notag\\
&\quad
-\frac{m^{3}}{36r^{7}}\left(
2268a^{4}\zeta^{4}-1512a^{4}\zeta^{2}+108a^{4}
-952a^{2}m^{2}\zeta^{2}+104a^{2}m^{2}+288m^{4}
\right)
+O(r^{-8}),\\
h_H^{01}
&=Gh_{\mathrm{can}(1)}^{01}
+n_2\Bigg[
\frac{2am^{2}}{r^{3}}
+\frac{2am^{3}}{r^{4}}
-\frac{2am^{2}\left(3a^{2}\zeta^{2}-a^{2}-m^{2}\right)}{r^{5}}
-\frac{am^{3}\left(7a^{2}\zeta^{2}-a^{2}-2m^{2}\right)}{r^{6}}
\notag\\
&\quad
+\frac{2am^{2}\left(
10a^{4}\zeta^{4}-8a^{4}\zeta^{2}+a^{4}
-4a^{2}m^{2}\zeta^{2}+a^{2}m^{2}+m^{4}
\right)}{r^{7}}
\Bigg]
+O(r^{-8}),\\
h_H^{11}
&=
-\frac{m^{2}n_1^{2}}{r^{2}}
+\frac{a^{2}m^{2}\left[
\zeta^{2}-1+(3\zeta^{2}+1)n_1^{2}
\right]}{r^{4}}
-\frac{a^{2}m^{2}(\zeta^{2}-1)
\left(12a^{2}\zeta^{2}-3a^{2}-4m^{2}\right)}{3r^{6}}
\notag\\
&\quad-\frac{a^{2}m^{2}
\left(30a^{2}\zeta^{4}-3a^{2}-5m^{2}\right)n_1^{2}}{3r^{6}}
+O(r^{-8}),\\
h_H^{12}
&=
-\frac{m^{2}n_1n_2}{r^{2}}
+\frac{a^{2}m^{2}(3\zeta^{2}+1)n_1n_2}{r^{4}}
-\frac{a^{2}m^{2}
\left(30a^{2}\zeta^{4}-3a^{2}-5m^{2}\right)n_1n_2}{3r^{6}}
+O(r^{-8}),\\
h_H^{13}
&=
+n_1\zeta\Bigg[
-\frac{m^{2}}{r^{2}}
+\frac{a^{2}m^{2}(3\zeta^{2}-1)}{r^{4}}
-\frac{a^{2}m^{2}
\left(30a^{2}\zeta^{4}-24a^{2}\zeta^{2}+3a^{2}-m^{2}\right)}
{3r^{6}}
\Bigg]
+O(r^{-8}),\\
h_H^{33}
&=
-\frac{m^{2}\zeta^{2}}{r^{2}}
+\frac{a^{2}m^{2}\zeta^{2}(3\zeta^{2}-2)}{r^{4}}
-\frac{a^{2}m^{2}\left[
90a^{2}\zeta^{6}-108a^{2}\zeta^{4}+27a^{2}\zeta^{2}
-3m^{2}\zeta^{2}+2m^{2}
\right]}{9r^{6}}
+O(r^{-8}).
\end{align}
\end{subequations}
\end{widetext}

\bibliographystyle{unsrt}
\bibliography{ref}
\end{document}